\documentclass[a4paper,aps,onecolumn]{revtex4}
\RequirePackage[english]{babel}
\RequirePackage[latin1]{inputenc}
\RequirePackage[T1]{fontenc}
\RequirePackage{mathrsfs}
\RequirePackage{amsmath}
\RequirePackage{amssymb}
\RequirePackage{amsbsy}
\RequirePackage{bm}
\def\de#1/de#2{\frac{\partial {#1}}{\partial {#2}}}
\begin{document}
\title{\textbf{Renormalizability of the Dirac Equation in Torsion-Gravity\\
with Non-Minimal Coupling}}
\author{Luca Fabbri$^{1}$\footnote{E-mail: fabbri@diptem.unige.it},
Stefano Vignolo$^{1}$\footnote{E-mail: vignolo@diptem.unige.it}, 
Sante Carloni$^{2}$\footnote{E-mail: sante.carloni@utf.mff.cuni.cz}}
\affiliation{$^{1}$DIME Sez. Metodi e Modelli Matematici, Universit\`{a} di Genova,\\
Piazzale Kennedy, Pad. D, 16129, Genova, ITALY\\
$^{2}$Institute of Theoretical Physics, Faculty of Mathematics and Physics\\ 
Charles University in Prague, V Hole\v{s}ovi\v{c}k\'{a}ch 2 180 00 Praha 8, CZECH REPUBLIC}
\date{\today}
\begin{abstract}
We will consider the torsional completion of gravity for a background filled with Dirac matter fields, studying what happens when fermionic non-minimal coupling is taken into account: we will show that, although non-minimal couplings are usually disregarded because of their ill-defined behaviour in ultraviolet regimes, this is due to the fact that torsion is commonly neglected, whereas when torsion is not left aside, even non-minimal couplings behave properly. In detail, we will see that non-minimal coupling allows to renormalize the Dirac equation even when torsion is taken into consideration and that in some type of non-minimally coupled models parity-oddness might be present even in the gravitational sector. In addition, we will show that in the presence of the considered non-minimal coupling, torsion is able to evade cosmological singularities as it can happen in the minimal coupling case and in some other non--minimally coupled theory. In the course of the paper, we shall consider a specific interaction as prototype to study this fermionic non-minimal coupling, but we will try to present results that do not depend on the actual structure of the non-minimal couplings by investigating alternative types of interaction.
\end{abstract}
\maketitle
\section{Introduction}
In physics, the construction of a given theory can be achieved once a Lagrangian is assigned: the infinity of theories that can be conceived corresponds to the infinity of Lagrangians that can be written, and quite astonishingly, once a field content is taken into account, Lorentz invariance and specific gauge symmetries are powerful tools to dramatically reduce the number of possible dynamics; then, additional requirements, such as renormalizability or conformal invariance, may set the final count to just a few. Remarkably, among these few are some of the best established theories we have ever had (the Dirac theory, the electro-weak theory, chromodynamics); but disappointingly, other very successful theories do not seem to fit in (for instance, the Einstein theory of gravitation is neither conformal nor renormalizable, its subsequent cosmology still lacks a proper explanation for inflation, a right place for dark matter, a complete understanding of dark energy). Enlarging the model as to have the possibility to host more physical phenomena demands for the corresponding relaxation of some requirements on the dynamical action and its Lagrangian.

In the present paper we will relax the constraint of minimal coupling. To better explain the situation, consider the fact that a Lagrangian is always given in terms of kinetic terms of given fields (that is, terms with one specific field) plus terms that describe the interactions among those fields (namely, terms with the product of more fields): because the field content we will consider here (torsion-gravity with Dirac fields) is rather well established, we are not going to change the structure of the kinetic terms; however, we will allow the interactions to be at the non-minimal mass dimension. The mass dimension of the interaction is given by assigning the mass dimension to each field (as it is known, in four-dimensional spacetimes, the mass dimension of derivatives and gauge potentials, as well as scalars, is $1$, for Dirac fermions is $\frac{3}{2}$, and so on): all kinetic terms are required to have mass dimension $4$, but for the interactions it is possible to have all terms of mass dimension $4$ (in which case the theory is conformally invariant) or all mass dimensions up to $4$ (in which case the theory is said to be renormalizable) or even all mass dimensions up to but excluding $4$ (in which case the theory is said to be super-renormalizable). It is customary not to consider mass dimensions larger than $4$ but in this paper considering larger mass dimensions is what we will do.

The reason for which larger mass dimensions, or equivalently non-minimal couplings, are in general not considered is that they violate renormalizability, that is the high-energy, or ultra-violet, behaviour is ill-defined; in these cases in fact, in the short-range scales these interactions will become dominant over the kinetic terms, and the theory would fail to be dynamical, according to Wilson criteria \cite{p-s}. 

In the present paper, we aim to generalize the work \cite{Carloni:2014pha} in the case torsion is not neglected, in order to provide a better understanding of the physics of non-minimal couplings: we will show that such phenomenology would not come at the expenses of the dynamics, because the non-minimal couplings we will consider will be proven to be renormalizable (i.e. dynamically well-defined). We will prove that such non-minimal couplings will even be super-renormalizable. Moreover, for consistency with the existing literature, we will show that the presented theory preserves the results on the avoidance of cosmological singularities, which have been already established
\cite{Poplawski:2009su,Fabbri:2011mg,Poplawski:2011jz,Magueijo:2012ug,Poplawski:2011wj,Poplawski:2010jv,Poplawski:2011qr}.

The paper will be divided in the following way: Section II is dedicated to the description of the theory and its general properties, particularly the conservation laws; Section III is instead dedicated to some interesting aspects of the theory such as renormalizability and parity violation; Section IV is devoted to cosmological considerations; Section V is dedicated to the summary of the results, and an overall discussion as conclusion.
\section{Torsion-Gravity with Dirac Fields in Non-Minimal Coupling}
To begin with, we shall introduce the general formalism, first defining the fundamental geometrical fields, then their dynamics with specific accent on non-minimal couplings; the last sub-section will be devoted to the proof that in the case of non-minimal couplings the conservation laws are improved with extra terms compared to the standard ones.

\subsection{The Geometrical Background}
In this paper, we indicate spacetime indices by Latin letters and Lorentz indices with Greek letters. The spacetime will be taken to be $(1\!+\!3)$-dimensional and endowed with a metric tensor $g_{ij}$ and a connection $\Gamma_{ij}^{\;\;\;h}$ which will be metric-compatible, that is the connection defines a covariant derivative $\nabla_{i}$ that once applied on the metric tensor yields zero; the connection defines Cartan torsion and Riemann curvature expressed in holonomic frames as
\begin{subequations}
\label{2.2}
\begin{equation}
\label{torsion}
T_{ij}^{\;\;\;h}=\Gamma_{ij}^{\;\;\;h}-\Gamma_{ji}^{\;\;\;h}
\end{equation}
\begin{equation}
\label{curvature}
R^{h}_{\;\;kij}=\partial_i\Gamma_{jk}^{\;\;\;h} - \partial_j\Gamma_{ik}^{\;\;\;h} +
\Gamma_{ip}^{\;\;\;h}\Gamma_{jk}^{\;\;\;p}-\Gamma_{jp}^{\;\;\;h}\Gamma_{ik}^{\;\;\;p}.
\end{equation}
\end{subequations}
The contractions $T_{i}=T_{ij}^{\;\;\;j}$, $R_{ij}=R^{h}_{\phantom{h}ihj}$ and $R=R_{ij}g^{ij}$ are called respectively the torsion vector, the Ricci tensor and the Ricci scalar curvature, and they verify the identities
\begin{subequations}
\label{JB}
\begin{equation}
\label{JB1}
\nabla_i\left(R^{ij} - \frac{1}{2}Rg^{ij}\right) = \frac{1}{2}T_{pqr}R^{pqrj} + T^{jpq}R_{pq}
\end{equation}
\begin{equation}
\label{JB2} 
\nabla_i\left(T^{tsi} - T^t g^{si} + T^s g^{ti}\right) = - T^{tsp}T_p - R^{ts} + R^{st}
\end{equation}
\end{subequations}
known as Bianchi identities. Given a metric tensor $g_{ij}$, every metric-compatible connection can be decomposed as
\begin{equation}
\label{2.3}
\Gamma_{ij}^{\;\;\;h}=\tilde{\Gamma}_{ij}^{\;\;\;h}-K_{ij}^{\;\;\;h}
\end{equation}
where 
\begin{equation}
\label{2.4}
K_{ij}^{\;\;\;h}
=\frac{1}{2}\left(-T_{ij}^{\;\;\;h}+T_{j\;\;\;i}^{\;\;h}-T^{h}_{\;\;ij}\right)
\end{equation}
is called the contorsion tensor and $\tilde{\Gamma}_{ij}^{\;\;\;h}$ is the Levi--Civita connection induced by the metric $g_{ij}$; the contorsion tensor satisfies $K_{i}^{\;\;jh}=-K_{i}^{\;\;hj}$ amounting to the metric-compatibility of \eqref{2.3}, and it has one contraction $K_{i}^{\;\;ij}=K^{j}$ from which we have also the identity $K_{i}=-T_{i}$. With the contorsion we can decompose the covariant derivative of the full connection $\nabla_{i}$ into covariant derivative of the Levi--Civita connection $\tilde{\nabla}_{i}$ plus contorsional contributions; in terms of this decomposition, we have the corresponding decomposition go the curvature as given according to
\begin{equation}
\label{2.5}
R_{ij}=\tilde{R}_{ij} + \tilde{\nabla}_jK_{hi}^{\;\;\;h} - \tilde{\nabla}_hK_{ji}^{\;\;\;h} + K_{ji}^{\;\;\;p}K_{hp}^{\;\;\;h} - K_{hi}^{\;\;\;p}K_{jp}^{\;\;\;h}
\end{equation}
where the Ricci curvature of the Levi--Civita connection is denoted by $\tilde{R}_{ij} $. In the next sections we shall consider gravity coupled with Dirac fields: as it is well known, the most suitable variables to describe fermion fields interacting with gravity are tetrads, which possess Lorentz indices as well as spacetime indices. They are defined together with their dual $e^{j}_{\mu}e^{\mu}_{i}=\delta^{j}_{i}$ and $e^{j}_{\mu}e^{\nu}_{j}=\delta^{\nu}_{\mu}$ in such a way that $g_{ij}\!=\!e^{\mu}_{i}e^{\nu}_{j}\eta_{\mu\nu}$ where $\eta_{\mu\nu}={\rm diag}(1,-1,-1,-1)$ is the Minkowskian matrix, and in addition we may also define the spin-connections $1$-forms $\omega^\mu_{\;\;\nu}=\omega_{i\;\;\;\nu}^{\;\;\mu}\,dx^i$ such that
\begin{equation}
\label{2.6}
\Gamma_{ij}^{\;\;\;h} = \omega_{i\;\;\;\nu}^{\;\;\mu}e_\mu^h e^\nu_j + e^{h}_{\mu}\partial_{i}e^{\mu}_{j}
\end{equation}
with $\omega_{i}^{\;\;\mu\nu}=-\omega_{i}^{\;\;\nu\mu}$ identically; torsion and curvature tensors can also be expressed as 
\begin{subequations}
\label{2.7}
\begin{equation}
\label{2.7a}
T^\mu_{ij}=\partial_i e^\mu_j-\partial_j e^\mu_i
+\omega^{\;\;\mu}_{i\;\;\;\lambda}e^\lambda_{j}-\omega^{\;\;\mu}_{j\;\;\;\lambda}e^\lambda_{i}
\end{equation}
\begin{equation}
\label{2.7b}
R_{ij}^{\;\;\;\;\mu\nu}
=\partial_i\omega_{j}^{\;\;\mu\nu} - \partial_j{\omega_{i}^{\;\;\mu\nu}}
+\omega^{\;\;\mu}_{i\;\;\;\lambda}\omega_{j}^{\;\;\lambda\nu}
-\omega^{\;\;\mu}_{j\;\;\;\lambda}\omega_{i}^{\;\;\lambda\nu}
\end{equation}
\end{subequations}
and their relationships with the world tensors defined in equations \eqref{2.2} are given by the relationships $T_{ij}^{\;\;\;h}=T^{\;\;\;\alpha}_{ij}e_\alpha^h$ and $R^{h}_{\;\;kij}
=R_{ij\;\;\;\;\nu}^{\;\;\;\;\mu}e_\mu^h e^\nu_k$ respectively. Spacetime and Lorentz formalisms are totally equivalent.

The spinor fields we will consider are basically the simplest possible, that is the $\frac{1}{2}$-spin spinor fields, or Dirac fields. They are introduced in terms of the Clifford algebra, which is constituted by the set of gamma matrices given by $\gamma^\mu$, and of course $\Gamma^i = e^i_\mu\gamma^\mu$, in terms of which we can define $S_{\mu\nu}= \frac{1}{8}[\gamma_\mu,\gamma_\nu]$ that can be proven to be the infinitesimal generators of a complex Lorentz transformations: the spinorial-covariant derivative of the spinor field are given by $D_i\psi=\de\psi/de{x^i}-\Omega_i\psi$ where the spinorial-connection is given according to
\begin{equation}\label{3.3}
\Omega_i=-\omega_i^{\;\;\mu\nu}S_{\mu\nu}\equiv-\frac{1}{4}g_{jh}\left(\Gamma_{ik}^{\;\;\;j}
-e^j_\mu\partial_i e^\mu_k \right)\Gamma^h\Gamma^k
\end{equation}
in terms of the spin-connection, or the linear connection. In the paper, we will consider no gauge interaction.

\subsection{The Dynamical Action}
With this field content, it possible to define the dynamical action: the kinematic sector will be given in terms of the usual action for the Einstein-Sciama-Kibble gravity filled with a Dirac field; however, as we anticipated in the Introduction, we will take into account also additional terms giving rise to interactions in a non-minimal fashion. The simplest terms we may think to add are those constituted of interactions of mass dimension $5$: because the interaction must contain geometrical as well as Dirac fields, and because Dirac fields are spinors that have to be bracketed with one another, then the spinorial contributions will already account for a mass dimension $\frac{3}{2}\cdot2=3$, and therefore the remaining $5-3=2$ mass dimensional term can only be a curvature or a squared-torsion or a derivative-torsion; the only term that would still be present even in the torsionless limit can only be a curvature, and as a quick algebra on the gamma matrices would show, all possible terms actually reduce to 
the single $R\bar\psi\psi$ so that 
\begin{equation}
\mathcal{L}_{\mathrm{interaction}}=[\epsilon R\bar\psi\psi-V(\psi\psi)]e
\end{equation}
with $\epsilon$ an undetermined parameter, will be our interacting term in the Lagrangian. The total Lagrangian is then
\begin{equation}
\mathcal{L}=[(1+\epsilon \bar\psi\psi)R-\frac{i}{2}(\bar\psi\Gamma^{i}D_{i}\psi-D_{i}\bar\psi\Gamma^{i}\psi)+m\psi\psi-V(\psi\psi)]e.
\end{equation}
Its variation with respect to the tetrad and spin-connection as well as Dirac fields gives
\begin{subequations}
\label{feq}
\begin{equation}
\label{feqT}
\varphi\left( T^{\;\;\;\alpha}_{ts} - T^{\;\;\;\sigma}_{t\sigma}e^\alpha_s + T^{\;\;\;\sigma}_{s\sigma}e^\alpha_t \right) =\de{\varphi}/de{x^t}e^\alpha_s - \de{\varphi}/de{x^s}e^\alpha_t + S^{\;\;\;\alpha}_{ts}
\end{equation}
\begin{equation}
\label{feqR}
R_{\mu\sigma}^{\;\;\;\;\lambda\sigma}e^{i}_\lambda -\frac{1}{2}Re^i_{\mu}=\frac{1}{\varphi}\Sigma^i_\mu 
\end{equation}
\end{subequations}
\begin{equation}
\label{feqM}
i\Gamma^{h}D_{h}\psi + \frac{i}{2}T_h\Gamma^h\psi- m\psi + V'(\bar\psi\psi)\psi - \epsilon\psi R=0
\end{equation}
where we have denoted $\varphi=(1+\epsilon\bar\psi\psi)$ and $V'=\frac{dV}{d(\bar\psi\psi)}$, for simplicity, together with the usual $\Sigma^i_\mu = - \frac{1}{2e}\frac{\partial(e{L}_D)}{\partial e^\mu_i}$ and $S^{\;\;\;\;i}_{\mu\nu}= \frac{1}{2e}\frac{\partial(e{L}_D)}{\partial\omega_i^{\;\;\mu\nu}}$ playing the role of energy and spin density tensors, respectively: they are given by expressions
\begin{subequations}
\label{conquan}
\begin{equation}
\label{spin}
S_{ijh}=\frac{i}{2}\bar\psi\left\{\Gamma^{h},S_{ij}\right\}\psi
\equiv-\frac{1}{4}\epsilon_{hijk}\bar{\psi}\Gamma_{5}\gamma^{k}\psi
\end{equation}
\begin{equation}
\label{energy}
\begin{split}
\Sigma_{ij}=\frac{i}{4} \left( \bar\psi\Gamma_{i}{D}_{j}\psi - {D}_{j}\bar{\psi}\Gamma_{i}\psi \right) - \frac{1}{2}\epsilon(\bar\psi\psi)R\,g_{ij} - \frac{1}{2}V(\bar\psi\psi)\,g_{ij} + \frac{1}{2}(\bar\psi\psi)V'(\bar\psi\psi)\,g_{ij}
\end{split}
\end{equation}
\end{subequations}
where \eqref{energy} is obtained by employing the Dirac equations \eqref{feqM}.

Using the set of field equations for torsion \eqref{feqT}, we obtain the following representation for the contorsion tensor
\begin{equation}\label{3.11}
K_{ij}^{\;\;\;h}= \hat{K}_{ij}^{\;\;\;h} + \hat{S}_{ij}^{\;\;\;h}
\end{equation}
where
\begin{subequations}\label{3.12}
\begin{equation}\label{3.12a}
\hat{S}_{ij}^{\;\;\;h}= - \frac{1}{2\varphi} S_{ij}^{\;\;\;h}
\end{equation}
\begin{equation}\label{3.12b}
\hat{K}_{ij}^{\;\;\;h} = -\hat{T}_j\delta^h_i + \hat{T}_pg^{ph}g_{ij}
\end{equation}
\begin{equation}\label{3.12c}
\hat{T}_j=\frac{1}{2\varphi}\de{\varphi}/de{x^j}
\end{equation}
\end{subequations}
and after splitting the spin into its three irreducible terms, only the two vector terms will remain, the completely antisymmetric part, sourced by the spin of the Dirac field, and the vector part, generated by the non-minimal coupling
\begin{eqnarray}
\label{aux}
T_{ak}^{\;\;\;k}=-\frac{3}{2\varphi}\partial_{a}\varphi.
\end{eqnarray}
Without the non-minimal coupling this term would vanish identically.

To proceed in our study, let us substitute \eqref{aux} into \eqref{feqT}, so to invert the expression for the torsion according to
\begin{equation}
\label{torinverted}
T_{tsa}=\frac{1}{2\varphi}\left(g_{at}\partial_{s}\varphi-g_{as}\partial_{t}\varphi\right)
+\frac{1}{4\varphi}\epsilon_{tsak}\bar{\psi}\Gamma^{k}\gamma_{5}\psi
\end{equation}
in which the role of the two vector parts is manifest; field equations \eqref{feqR} are symmetrized and decomposed as
\begin{equation}
\label{curvature-energy}
\begin{split}
\tilde{R}_{ij}-\frac{1}{2}\tilde{R}g_{ij}=-\frac{1}{2\varphi}\epsilon \bar\psi\psi R\,g_{ij}
+\frac{1}{\varphi}\tilde{\Sigma}_{ij}+\frac{1}{\varphi^2}\left( - \frac{3}{2}\de\varphi/de{x^i}\de\varphi/de{x^j} + \varphi\tilde{\nabla}_{j}\de\varphi/de{x^i} + 
\frac{3}{4}\de\varphi/de{x^h}\de\varphi/de{x^k}g^{hk}g_{ij}- \varphi\tilde{\nabla}^h\de\varphi/de{x^h}g_{ij}\right)+\\
+\frac{3}{64\varphi^2}(\bar{\psi}\gamma_5\gamma^\tau\psi)(\bar{\psi}\gamma_5\gamma_\tau\psi)g_{ij}
-\frac{1}{2\varphi}V(\bar\psi\psi)\,g_{ij}+\frac{1}{2\varphi}(\bar\psi\psi)V'(\bar\psi\psi)\,g_{ij}
\end{split}
\end{equation}
with
\begin{equation}\label{3.15}
\tilde{\Sigma}_{ij} = \frac{i}{4} \left[ \bar\psi\Gamma_{(i}\tilde{D}_{j)}\psi -\tilde{D}_{(j}\bar\psi\Gamma_{i)}\psi \right]
\end{equation}
where $\tilde{D}_{j}\psi$ is the spinorial covariant derivative with respect to the Levi-Civita connection. The antisymmetric part of \eqref{feqR} is simply the conservation law for the spin, ensured by the Dirac equations \cite{VFC}, as we are going to show in the following section. Dirac equations \eqref{feqM} are given according to the expression
\begin{equation}
\label{mattereq}
i\Gamma^{h}\tilde{D}_{h}\psi-\frac{3}{16\varphi}\left[(\bar{\psi}\psi)
+i(i\bar{\psi}\gamma_5\psi)\gamma_5\right]\psi
-\left(m-V'(\bar\psi\psi)+\epsilon R\right)\psi=0
\end{equation}
as it can be checked by following the procedure outlined in \cite{FV1}.

These equations, however, have a problem: if we separate the interacting term in its torsionless term $\tilde{R}$ and torsional contributions, and we substitute torsion with the matter field, \eqref{mattereq} becomes
\begin{equation}
\begin{split}
\frac{3\epsilon^{2}}{\varphi}\tilde{\nabla}^{2}(\bar\psi\psi)\psi
-\frac{3\epsilon^{3}}{2\varphi^{2}}\tilde{\nabla}_i(\bar\psi\psi)\tilde{\nabla}^i(\bar\psi\psi)\psi
+i\Gamma^{h}\tilde{D}_{h}\psi+\frac{3\epsilon}{32\varphi^{2}}\left[(\bar{\psi}\psi)^{2}
+(i\bar{\psi}\gamma_5\psi)^{2}\right]\psi+\\
-\frac{3}{16\varphi}\left[(\bar{\psi}\psi)
+i(i\bar{\psi}\gamma_5\psi)\gamma_5\right]\psi
-\left(m-V'(\bar\psi\psi)+\epsilon \tilde{R}\right)\psi=0
\end{split}
\end{equation}
containing second-order time derivatives of the spinor that cannot be present into a Dirac field equations. Indeed, it is possible to prove that when the field equations contain second-order time derivatives of the spinor, there is the possibility to induce a mismatch of number of degrees of freedom and violation of causal propagation, as it has been described in terms of the Velo-Zwanziger analysis \cite{v-z/1,v-z/2}. Nevertheless, things can be straightened up by employing the coupling to gravity. To see that, consider that the trace of the gravitational field equations yields
\begin{equation}\label{3.16}
-\varphi R=\Sigma^{i}_{i}=\frac{m}{2}\bar\psi\psi-\frac{3}{2}\epsilon\bar\psi\psi R-2V
+\frac{3}{2}\bar\psi\psi V'
\end{equation}
made explicit after inverting it as
\begin{equation}\label{3.18}
R=\frac{\left(\frac{m}{2}\bar\psi\psi-2V+\frac{3}{2}\bar\psi\psi V'\right)}{\left(\frac{1}{2}\varphi-\frac{3}{2}\right)}
\end{equation}
relating the Ricci scalar to the bilinear spinors, that is linking the two additional interacting terms we have decided to study. When the equation \eqref{3.18} is substituted back into the gravitational field equations, they become
\begin{equation}\label{3.19}
\begin{split}
\tilde{R}_{ij}-\frac{1}{2}\tilde{R}g_{ij}=\frac{i}{4\varphi} \left[ \bar\psi\Gamma_{(i}\tilde{D}_{j)}\psi-\tilde{D}_{(j}\bar\psi\Gamma_{i)}\psi \right]+
\frac{1}{\varphi^2}\left(-\frac{3}{2}\de\varphi/de{x^i}\de\varphi/de{x^j} + \varphi\tilde{\nabla}_{j}\de\varphi/de{x^i} + 
\frac{3}{4}\de\varphi/de{x^h}\de\varphi/de{x^k}g^{hk}g_{ij}- \varphi\tilde{\nabla}^h\de\varphi/de{x^h}g_{ij}\right)+\\
+\frac{3}{64\varphi^2}(\bar{\psi}\gamma_5\gamma^\tau\psi)(\bar{\psi}\gamma_5\gamma_\tau\psi)g_{ij}
-\frac{1}{2\varphi}V(\bar\psi\psi)\,g_{ij}+\frac{1}{2\varphi}(\bar\psi\psi)V'(\bar\psi\psi)\,g_{ij}-\frac{\epsilon\bar\psi\psi\left(\frac{m}{2}\bar\psi\psi -2V + \frac{3}{2}\bar\psi\psi V'\right)}{2\varphi\left(\frac{1}{2}\varphi - \frac{3}{2}\right)}g_{ij}.
\end{split}
\end{equation}
The substitution into the Dirac field equation instead gives
\begin{equation}
\begin{split}
i\Gamma^{h}\tilde{D}_{h}\psi-\frac{3}{16\varphi}\left[(\bar{\psi}\psi)
+i(i\bar{\psi}\gamma_5\psi)\gamma_5\right]\psi
-\left(\frac{-m+V'(\bar\psi\psi)+m\epsilon\bar\psi\psi-2\epsilon V+\epsilon\bar\psi\psi V'}{\frac{1}{2}\epsilon\bar\psi\psi-1}\right)\psi=0 
\end{split}
\end{equation}
showing now no sign of second-order time derivative, nor first-order time derivative apart from the usual Dirac operator, but only the expected cubic term plus a new term containing the interaction with gravity encoded by the parameter $\epsilon$ and the potential of self-interaction $V$ alone, and so causality is preserved \cite{v-z/1,v-z/2}.

Clearly, a judicious choice of the potential $V$ could give rise to correspondingly fine-tuned behaviours, but of particular interest is the special case in which the potential of self-interaction is absent, giving
\begin{equation}
\begin{split}
\label{Einstein}
\tilde{R}_{ij}-\frac{1}{2}\tilde{R}g_{ij}=\frac{1}{\varphi}\tilde{\Sigma}_{ij}
+\frac{1}{\varphi^2}\left( - \frac{3}{2}\de\varphi/de{x^i}\de\varphi/de{x^j} + \varphi\tilde{\nabla}_{j}\de\varphi/de{x^i} + 
\frac{3}{4}\de\varphi/de{x^h}\de\varphi/de{x^k}g^{hk}g_{ij}- \varphi\tilde{\nabla}^h\de\varphi/de{x^h}g_{ij}\right)+\\
-\left[\frac{3}{64\varphi^2}|(\bar{\psi}\psi)^{2}+(i\bar{\psi}\gamma_5\psi)^{2}|
+\frac{m\epsilon\bar\psi\psi\bar\psi\psi}{2\varphi\left(\varphi-3\right)}\right]g_{ij}
\end{split}
\end{equation}
and
\begin{equation}
\label{Dirac}
i\Gamma^{h}\tilde{D}_{h}\psi-m\left[\frac{3}{16m\varphi}\left[(\bar{\psi}\psi)
+i(i\bar{\psi}\gamma_5\psi)\gamma_5\right]
+\left(\frac{\epsilon\bar\psi\psi-1}{\frac{1}{2}\epsilon\bar\psi\psi-1}\right)\right]\psi=0 
\end{equation}
in which we see that the new interaction with gravity has the same influence as the torsionally-induced non-linearities of the matter field. This remark will prove to be fundamental when discussing the issue of renormalizability.

\subsection{Conservation laws}
Having now the fundamental field equations, we will employ the contracted Bianchi identities in order to derive the conservation laws for the theory outlined above. The derivation of these conservation laws is not a mere exercise, but it is essential to show the validity of conservation laws that are no longer given by the usual ones, but which are instead improved by the presence of terms that due their existence to the extra non-minimal interaction; the usual form of conservation laws (as proven in reference \cite{h-h-k-n}, for example) can be obtained from underlying symmetries by applying N\"{o}ther's theorem, but this has never been done while taking into account the non-minimal coupling mixing gravitational and material sectors: and the extra terms of the modified conservation laws come from those terms.

Let us consider the fully contracted Bianchi identities given according to \eqref{JB} and the field equations \eqref{feqR} as
\begin{subequations}\label{c.2.2}
\begin{equation}\label{c.2.2a}
\varphi \left(R^{ij} -\frac{1}{2} R g^{ij}\right)= \Sigma^{ij}
\end{equation}
\begin{equation}\label{c.2.2b}
\varphi \left(T^{ijh} -T^i g^{jh} + T^j g^{ih}\right)= \nabla^i\varphi g^{jh} - \nabla^j\varphi g^{ih} + S^{ijh}
\end{equation}
\end{subequations} 
so that through the divergences of \eqref{c.2.2} we get 
\begin{subequations}\label{c.2.3}
\begin{equation}\label{c.2.3a}
\nabla_i\varphi \left(R^{ij} -\frac{1}{2} R g^{ij}\right) + \varphi\nabla_i\left(R^{ij} -\frac{1}{2} R g^{ij}\right) = \nabla_i\Sigma^{ij} 
\end{equation}
\begin{equation}\label{c.2.3b}
\nabla_h\varphi\left(T^{ijh} -T^i g^{jh} + T^j g^{ih}\right) + \varphi\nabla_h\left(T^{ijh} -T^i g^{jh} + T^j g^{ih}\right) = \left[\nabla^j,\nabla^i\right]\varphi + \nabla_h S^{ijh}.
\end{equation}
\end{subequations}
By inserting the content of \eqref{JB} into \eqref{c.2.3} and evaluating explicitly $\left[\nabla^j,\nabla^i\right]\varphi $ we obtain
\begin{subequations}\label{c.2.4}
\begin{equation}\label{c.2.4a}
\nabla_i\varphi \left(R^{ij} -\frac{1}{2} R g^{ij}\right) + \frac{1}{2}\varphi T_{pqr}R^{pqrj} + \varphi T^{jpq}R_{pq} = \nabla_i\Sigma^{ij} 
\end{equation}
\begin{equation}\label{c.2.4b}
\nabla^i\varphi T^j - \nabla^j\varphi T^i - \varphi T^{ijp}T_p - \varphi R^{ij} + \varphi R^{ji} = \nabla_h S^{ijh}.
\end{equation}
\end{subequations}
Moreover, from the antisymmetric part of \eqref{c.2.2a} we have
\begin{subequations}\label{c.2.5}
\begin{equation}\label{c.2.5a}
\varphi \left(R^{ij} - R^{ji}\right) = \Sigma^{ij} - \Sigma^{ji}
\end{equation}
and the contractions of \eqref{c.2.2a} with torsion vector and torsion yield
\begin{equation}\label{c.2.5b}
\varphi R^{ij}T_i - \frac{1}{2}\varphi R T^j = \Sigma^{ij}T_i 
\end{equation}
\begin{equation}\label{c.2.5c}
\varphi R^{ij}T_{hij} - \frac{1}{2}\varphi R T_h = \Sigma^{ij}T_{hij} 
\end{equation}
as well as the contractions of \eqref{c.2.2b} with curvature and torsion vector give rise to
\begin{equation}\label{c.2.5d}
\frac{1}{2}\varphi T^{ijh}R_{ijhk} + \varphi T^i R_{ik} = - \nabla^i\varphi R_{ik} + \frac{1}{2}S^{ijh}R_{ijhk}
\end{equation}
\begin{equation}\label{c.2.5e}
\varphi T^{ijh}T_h = \nabla^i\varphi T^j - \nabla^j\varphi T^i + S^{ijh}T_h
\end{equation}
\end{subequations}
as a final step. Eventually, by inserting \eqref{c.2.5} into \eqref{c.2.4}, we end up with the conservation laws
\begin{subequations}
\label{cons}
\begin{equation}
\label{consenergy}
\nabla_i\Sigma^{ij} + T_i\Sigma^{ij} - \Sigma_{pq}T^{jpq} -\frac{1}{2}S_{pqr}R^{pqrj} + \frac{1}{2}R\nabla^j \varphi= 0
\end{equation}
\begin{equation}
\label{consspin}
\nabla_h S^{ijh} + T_h S^{ijh} + \Sigma^{ij} - \Sigma^{ji} = 0 
\end{equation}
\end{subequations}
which the energy--momentum \eqref{energy} and spin \eqref{spin} tensors have to satisfy. We see that equation \eqref{energy} is different from the standard one because of the presence of the last term $R\nabla^j \varphi$ which arises due to the coupling of the condensate and geometry, that is the non-minimal coupling we are studying in the present context.

For completeness and consistency, we verify that \eqref{feqM} imply \eqref{cons}. We start decomposing the energy tensor as
\begin{equation}\label{c.2.7}
\Sigma_{ij} = \bar{\Sigma}_{ij} + \hat{\Sigma}_{ij}
\end{equation}
where
\begin{equation}\label{c.2.7a}
\bar{\Sigma}_{ij} = \frac{i}{4} \left( \bar\psi\Gamma_{i}{D}_{j}\psi - {D}_{j}\bar{\psi}\Gamma_{i}\psi \right)
\end{equation}
and
\begin{equation}\label{c.2.7b}
\hat{\Sigma}_{ij}=-\frac{1}{2}\epsilon(\bar\psi\psi)Rg_{ij}
-\frac{1}{2}V(\bar\psi\psi)\,g_{ij} + \frac{1}{2}(\bar\psi\psi)V'(\bar\psi\psi)\,g_{ij}
\end{equation}
as the purely kinetic and potential parts. On the one hand making use of \eqref{2.3}, \eqref{2.4} and taking the symmetry of $\hat{\Sigma}_{ij}$ into account, it is a straightforward matter to verify the identities
\begin{eqnarray}\label{c.2.8}
\nonumber
&\nabla_i\hat{\Sigma}^{ij} = - T_i\hat{\Sigma}^{ij} + \hat{\Sigma}_{pq}T^{jpq} + \tilde{\nabla}_i\hat{\Sigma}^{ij} =\\
&=- T_i\hat{\Sigma}^{ij} + \hat{\Sigma}_{pq}T^{jpq} - \frac{1}{2}\tilde{\nabla}^j\left(\epsilon\bar\psi\psi\right)R - \frac{1}{2}\epsilon\bar\psi\psi\tilde{\nabla}^j R + \frac{1}{2}\left[\tilde{\nabla}^j V'\left(\bar\psi\psi\right)\right]\left(\bar\psi\psi\right)
\end{eqnarray}
while on the other hand, we can calculate the divergence of $\bar{\Sigma}_{ij} $ as
\begin{equation}\label{c.2.9}
D_{i}\bar{\Sigma}^{ij}
=\frac{i}{4}\left(D_{i}\bar\psi\Gamma^{i}D^{j}\psi+\bar\psi\Gamma^{i}D_{i}D^{j}\psi
-D_{i}D^{j}\bar\psi\Gamma^{i}\psi-D^{j}\bar\psi\Gamma^{i}D_{i}\psi\right)
\end{equation}
so that by adding and removing the terms $\bar{\psi}\Gamma^iD^jD_i\psi$ and $D^jD_i\bar{\psi}\Gamma^i\psi$ in equations \eqref{c.2.9} we get 
\begin{eqnarray}\label{c.2.10}
\nonumber
&D_{i}\bar{\Sigma}^{ij}
=\frac{i}{4}(\bar\psi\Gamma_{i}[D^{i},D^{j}]\psi-[D^{i},D^{j}]\bar\psi\Gamma_{i}\psi+\\
&+D_{i}\bar\psi\Gamma^{i}D^{j}\psi+\bar\psi\Gamma_{i}D^{j}D^{i}\psi
-D^{j}D^{i}\bar\psi\Gamma_{i}\psi-D^{j}\bar\psi\Gamma^{i}D_{i}\psi)
\end{eqnarray}
in which the commutator of covariant derivatives can be computed in terms of curvature and torsion terms; by inserting the Dirac equations \eqref{feqM} into \eqref{c.2.10} we obtain
\begin{equation}\label{c.2.12}
D_{i}\bar{\Sigma}^{ij}=- T_i\bar{\Sigma}^{ij} + T^{jik}\bar{\Sigma}_{ik}+\frac{1}{2}S_{abi}R^{abij} - \frac{1}{2}\left[\tilde{\nabla}^j V'\left(\bar\psi\psi\right)\right]\left(\bar\psi\psi\right) + \frac{1}{2}\epsilon\bar\psi\psi\tilde{\nabla}^j R.
\end{equation}
Adding \eqref{c.2.8} and \eqref{c.2.12} gives \eqref{consenergy}. As to the conservation law for the spin, we first calculate the divergence
\begin{equation}\label{c.2.14}
D_{h}S^{ijh}
=\frac{i}{2}\left(D_{h}\bar\psi\Gamma^{h}S^{ij}\psi+\bar\psi\Gamma^{h}S^{ij}D_{h}\psi
+D_{h}\bar\psi S^{ij}\Gamma^{h}\psi+\bar\psi S^{ij}\Gamma^{h}D_{h}\psi\right)
\end{equation}
and then we sum and subtract the terms $\bar{\psi}S^{ij}\gamma^hD_h\psi$ and $D_h\bar{\psi}\gamma^hS^{ij}\psi$, obtaining
\begin{equation}\label{c.2.15}
D_{h}S^{ijh}
=\frac{i}{2}(\bar\psi[\Gamma^{h},S^{ij}]D_{h}\psi+D_{h}\bar\psi [S^{ij},\Gamma^{h}]\psi
+2D_{h}\bar\psi\Gamma^{h}S^{ij}\psi+2\bar\psi S^{ij}\Gamma^{h}D_{h}\psi)
\end{equation}
where the commutator of gamma and sigma matrices can be evaluated in terms of their algebraic relationships; upon insertion of the Dirac equations \eqref{feqM} into \eqref{c.2.15} we get
\begin{equation}\label{c.2.16}
D_{h}S^{ijh}
=\frac{i}{2}(\bar\psi[\Gamma^{h},S^{ij}]D_{h}\psi+D_{h}\bar\psi [S^{ij},\Gamma^{h}]\psi - T_h\bar{\psi}\{\Gamma^h,S^{ij}\}\psi)
\end{equation}
similarly as above: finally, using
$[\Gamma^{i},[\Gamma^{j},\Gamma^{h}]]=4\left(\Gamma^{h}g^{ij}-\Gamma^{j}g^{ih}\right)$ as well as of the symmetry of $\hat{\Sigma}_{ij}$, we derive \eqref{consspin}.

For later use, we now show that the conservation laws \eqref{cons} (and then the Dirac equations \eqref{feqM}) ensure that the Levi--Civita divergence of the symmetrized Einstein--like equations \eqref{curvature-energy} vanishes. To start, recall that 
\begin{equation}\label{4.A.2}
R^h_{\;\;iqj}= \tilde{R}^h_{\;\;iqj}+
\tilde{\nabla}_jK_{qi}^{\;\;\;h}-\tilde{\nabla}_qK_{ji}^{\;\;\;h}+
K_{ji}^{\;\;\;p}K_{qp}^{\;\;\;h}-K_{qi}^{\;\;\;p}K_{jp}^{\;\;\;h}
\end{equation}
where the contorsion tensor $K_{ij}^{\;\;\;h}$ is expressed as in \eqref{2.4}. Taking \eqref{3.11} and \eqref{3.12} into account gives
\begin{equation}\label{4.A.5}
-\frac{1}{2}S_{hiq}R^{hiqj} = - S_h^{\;\;ip}\hat{S}^j_{\;\;ip}\hat{T}^h - \frac{1}{2}\varphi\tilde\nabla^j \left(\hat{S}_{qih}\hat{S}^{qih}\right) - \varphi\tilde\nabla_i \left(\hat{S}^{hqi}\hat{S}^j_{\;\;qh}\right) + \varphi\tilde\nabla_i \left(\hat{S}^{hqi}\right) \hat{S}^j_{\;\;qh} 
\end{equation}
as an identity; also, making use again of \eqref{3.11} and \eqref{3.12}, it is seen that
\begin{equation}\label{4.A.6}
\nabla_{i}\Sigma^{ij}+T_{i}\Sigma^{ij}-\Sigma_{ih}T^{jih} = \tilde\nabla_{i}\bar{\Sigma}^{(ij)} + \tilde\nabla_{i}\bar{\Sigma}^{[ij]} + \tilde{\nabla}_i\hat{\Sigma}^{ij} - K_{ih}^{\;\;\;j}\bar{\Sigma}^{[ih]} - T^j_{\;\;ih}\bar{\Sigma}^{[ih]}
\end{equation}
and moreover, it is an easy matter to verify that $\bar{\Sigma}^{(ij)} = \tilde{\Sigma}^{ij} - \varphi\hat{S}^{hip}\hat{S}^j_{\;\;ph}$, as well as that the conservation laws \eqref{consspin} amount to the identities $\frac{1}{\varphi}\bar{\Sigma}^{[ij]}+\tilde\nabla_h\hat{S}^{jih}=0$, which is the antisymmetrized part of the Einstein--like equations. From this identity and relationships \eqref{3.11} and \eqref{3.12} we deduce the following relations
\begin{subequations}\label{4.A.7}
\begin{equation}\label{4.A.7a}
\tilde\nabla_i\bar{\Sigma}^{(ij)} = \tilde\nabla_i\tilde{\Sigma}^{ij} - \tilde\nabla_i \left(\varphi\hat{S}^{hip}\hat{S}^{j}_{\;\;ph}\right)
\end{equation}
\begin{equation}\label{4.A.7b}
\tilde\nabla_i\bar{\Sigma}^{[ih]} = - \varphi_i\tilde\nabla_h\hat{S}^{jih}
\end{equation}
\begin{equation}\label{4.A.7c}
- K_{ih}^{\;\;\;j}\bar{\Sigma}^{[ih]} = - \varphi \hat{T}_h\tilde\nabla_q\hat{S}^{hjq} + \varphi\hat{S}_{ih}^{\;\;\;j}\tilde\nabla_q\hat{S}^{hiq}
\end{equation}
\begin{equation}\label{4.A.7d}
- T^j_{\;\;ih}\bar{\Sigma}^{[ih]} = \varphi \hat{T}_i\tilde\nabla_q\hat{S}^{jiq} - 2\varphi\hat{S}_{jih}\tilde\nabla_q\hat{S}^{hiq}
\end{equation}
\end{subequations}
where for simplicity we have defined $\varphi_i = \de\varphi/de{x^i}$. Inserting \eqref{4.A.5}, \eqref{4.A.6} and \eqref{4.A.7} into the conservation law for the energy \eqref{consenergy}, the latter can be expressed in the equivalent form given by the following expression
\begin{equation}\label{4.A.8}
\tilde\nabla_i\tilde{\Sigma}^{ij} + \tilde{\nabla}_i\hat{\Sigma}^{ij} - \frac{1}{2}\varphi\tilde\nabla^j\left(\hat{S}_{hqp}\hat{S}^{hqp}\right) + \frac{1}{2}R\tilde{\nabla}^j\varphi=0.
\end{equation}
In addition to this, from \eqref{2.5} and again \eqref{3.11} and \eqref{3.12} we derive the identity
\begin{equation}\label{4.A.11}
R = \tilde{R} + \frac{3}{2}\frac{1}{\varphi^2}\varphi_i\varphi^i - \frac{3}{\varphi}\tilde{\nabla}_i\varphi^i - \hat{S}_{hqp}\hat{S}^{hqp}
\end{equation}
and we write the symmetrized and decomposed Einstein--like equations in the form
\begin{equation}\label{4.A.12}
\begin{split}
\varphi\tilde{R}_{ij} -\frac{\varphi}{2}\tilde{R}g_{ij}= \tilde\Sigma_{ij} + \hat{\Sigma}_{ij} + \frac{1}{\varphi}\left( - \frac{3}{2}\varphi_i\varphi_j + \varphi\tilde{\nabla}_{j}\varphi_i + \frac{3}{4}\varphi_h\varphi^h g_{ij} + - \varphi\tilde{\nabla}^h\varphi_h g_{ij}\right)
-\frac{1}{2}\varphi\hat{S}_{hqp}\hat{S}^{hqp}g_{ij}.
\end{split}
\end{equation}
The Levi--Civita covariant divergence of \eqref{4.A.12} is
\begin{equation}\label{4.A.13}
\begin{split}
(\tilde\nabla^j\varphi)\tilde{R}_{ij} + \varphi\tilde\nabla^j\tilde{G}_{ij} -\frac{1}{2}\tilde{R}\tilde\nabla_i\varphi = \tilde\nabla^j\tilde\Sigma_{ij} + \tilde{\nabla}^j\hat{\Sigma}_{ij} + \left(\tilde\nabla^j\tilde{\nabla}_{j}\tilde\nabla_i - \tilde\nabla_i\tilde{\nabla}^j\tilde\nabla_j\right)\varphi +\\
+ \tilde\nabla^j\left[\frac{1}{\varphi}\left(-\frac{3}{2}\varphi_i\varphi_j + \frac{3}{4}\varphi_h\varphi^h g_{ij}\right)\right] - \frac{1}{2}\tilde\nabla_i\left(\varphi\hat{S}_{hqp}\hat{S}^{hqp}\right)
\end{split}
\end{equation}
and the fact that $\tilde\nabla^j\tilde{G}_{ij}=0$ and $(\tilde\nabla^j\varphi)\tilde{R}_{ij} = \left(\tilde\nabla^j\tilde{\nabla}_{j}\tilde\nabla_i - \tilde\nabla_i\tilde{\nabla}^j\tilde\nabla_j\right)\varphi$ reduces \eqref{4.A.13} to
\begin{equation}\label{4.A.14}
-\frac{1}{2}\tilde{R}\tilde\nabla_i\varphi = \tilde\nabla^j\tilde\Sigma_{ij} + \tilde{\nabla}^j\hat{\Sigma}_{ij} + \tilde\nabla^j\left[\frac{1}{\varphi}\left(-\frac{3}{2}\varphi_i\varphi_j + \frac{3}{4}\varphi_h\varphi^h g_{ij}\right)\right] - \frac{1}{2}\tilde\nabla_i\left(\varphi\hat{S}_{hqp}\hat{S}^{hqp}\right).
\end{equation}
From \eqref{4.A.11} it is seen that
\begin{equation}\label{4.A.15}
\frac{1}{2}R\varphi_i = +\frac{1}{2}\tilde{R}\varphi_i + \tilde\nabla^j\left[\frac{1}{\varphi}\left(-\frac{3}{2}\varphi_i\varphi_j + \frac{3}{4}\varphi_h\varphi^h g_{ij}\right)\right] - \frac{1}{2}\varphi_i\hat{S}_{hqp}\hat{S}^{hqp}
\end{equation}
and then \eqref{4.A.14} amount to the decomposed conservation laws \eqref{4.A.8}.
\section{The Effects of Non-Minimal Coupling: recovering renormalizability and parity violation}
Now that the non-minimal dynamics has been defined, we will investigate some consequences: as with torsion there are many non-minimal terms, even if we restrict ourselves to the simplest non-minimal couplings, our general behaviour will be that of focusing on the previous one as prototype, since we believe it to embody the most important properties; nevertheless, later on we will also switch to other non-minimal couplings, to show that indeed the features we will investigate are general aspects of the non-minimally coupled models, and not only for the simplest of them.
\subsection{Recovering Renormalizability}
As well known, the Dirac equation is non-renormalizable if torsion is present, in the usual situation of non-minimal coupling; in parallel, it is also well known that such an equation is non-renormalizable even without torsion, whenever non-minimal couplings are accounted for: given these premises, one would expect that the non-renormalizability would be much worsened if both torsion and non-minimal coupling are assumed. Contrary to this intuition, however, it will turn out that the Dirac equation with torsion \emph{and} non-minimal coupling is renormalizable; in fact, it is even super-renormalizable. As one of the reasons for which torsion had always been neglected was precisely the alleged non-renormalizability of the resulting Dirac equations, this result is fundamental because the renormalizability of the Dirac equation does not imposes us to neglect torsion, but it merely suggests that more non-minimal couplings might have to be considered, when dealing with Dirac fields in a gravitational underlying background.

So, to show that with both torsion and non-minimal coupling, the Dirac field is renormalizable, we will first focus on the case of no self-interaction of the spinor field, given when the potential $V$ is absent: \eqref{Einstein} and \eqref{Dirac} reduce to
\begin{equation}
\begin{split}
\tilde{R}_{ij}-\frac{1}{2}\tilde{R}g_{ij}=\frac{1}{\varphi}\tilde{\Sigma}_{ij}
+\frac{1}{\varphi^2}\left( - \frac{3}{2}\de\varphi/de{x^i}\de\varphi/de{x^j} + \varphi\tilde{\nabla}_{j}\de\varphi/de{x^i} + 
\frac{3}{4}\de\varphi/de{x^h}\de\varphi/de{x^k}g^{hk}g_{ij}- \varphi\tilde{\nabla}^h\de\varphi/de{x^h}g_{ij}\right)-\\
-\left[\frac{3}{64\varphi^2}|(\bar{\psi}\psi)^{2}+(i\bar{\psi}\gamma_5\psi)^{2}|
+\frac{m\epsilon\bar\psi\psi\bar\psi\psi}{2\varphi\left(\varphi-3\right)}\right]g_{ij}
\end{split}
\end{equation}
and
\begin{equation}
i\Gamma^{h}\tilde{D}_{h}\psi-m\left[\frac{3}{16m\varphi}\left[(\bar{\psi}\psi)
+i(i\bar{\psi}\gamma_5\psi)\gamma_5\right]
+\left(\frac{\epsilon\bar\psi\psi-1}{\frac{1}{2}\epsilon\bar\psi\psi-1}\right)\right]\psi=0 
\label{1}
\end{equation}
in which the non-minimal coupling and the torsionally-induced terms have the same influence on the matter field.

Focusing on the matter field equations, we consider Wilson's analysis of renormalizability: the general idea is that of assigning a scaling transformation property to the fundamental fields, checking the behaviour at very short distances. If the kinetic term becomes negligible with respect to the interaction, the equation is non-renormalizable; if the kinetic term survives, the equation is renormalizable, and in particular if it survives alone dominating all interactions, the equation is super-renormalizable \cite{p-s}. In our case, we only have spinors, which scale according to $\sigma^{-\frac{3}{2}}$; correspondingly
\begin{equation}
\begin{split}
&0=i\Gamma^{h}\tilde{D}_{h}\psi
-\frac{3m\sigma^{-1}}{16m(1+\epsilon\bar\psi\psi\sigma^{-3})}\left[(\bar{\psi}\psi)
+i(i\bar{\psi}\gamma_5\psi)\gamma_5\right]\psi
-\left(\frac{2\epsilon\bar\psi\psi\sigma^{-3}-2}{\epsilon\bar\psi\psi\sigma^{-3}-2}\right)m\sigma^{2}\psi\approx\\
&\approx i\Gamma^{h}\tilde{D}_{h}\psi
-\sigma^{2}\left(\frac{3}{16\epsilon\bar\psi\psi}\left[(\bar{\psi}\psi)
+i(i\bar{\psi}\gamma_5\psi)\gamma_5\right]-2m\right)\psi\approx i\Gamma^{h}\tilde{D}_{h}\psi
\end{split}
\end{equation}
implying that the non-minimal coupling and the torsionally-induced non-linear interactions work together to render the interaction, like the mass term, renormalizable. Or to be even more precise, super-renormalizable.

This is an important feature, and in order to avoid the thought that such a property could be ascribed to the specific form of the action we have been considering up to now, we will repeat the analysis in terms of another, simpler action.

Let us consider, for instance, a non-minimal coupling arising as a direct product between bilinear spinors and torsion terms, for which in general there are up to $20$ such interactions. However, since we are dealing with a toy-model to exemplify the main idea, there will be no great loss in choosing a case special enough to simplify computation. 

In addition, since among the three irreducible decompositions of torsion the completely antisymmetric one has a peculiar significance \cite{xy,a-l,m-l,Fabbri:2006xq,Fabbri:2009se}, we will pick torsion to be completely antisymmetric for the moment. Therefore we restrict torsion to be given by $T_{ijh}=\epsilon_{ijhk}W^{k}$ in terms of the axial vector $W^{k}$, so that the total number of possible terms reduces down to the three $\bar\psi\psi W^{k}W_{k}$, $i\bar\psi\gamma^5\psi \nabla_{k}W^{k}$, $\bar\psi\gamma^5S^{ij}\psi \nabla_{i}W_{j}$ alone, and the torsional interaction Lagrangian is
\begin{equation}
\mathcal{L}_{\mathrm{interaction}}=(A\bar\psi\psi W^{k}W_{k}+iB\bar\psi\gamma^5\psi \nabla_{k}W^{k}
+C\bar\psi\gamma^5S^{ij}\psi \nabla_{i}W_{j})\sqrt{|g|}
\end{equation}
in terms of three parameters $A$, $B$, $C$ only, which will be further reduced.

When all these terms are taken in the total action with matter, we have that variation with respect to all torsion or spin-connection and the spinor field gives the torsion-spin field equations
\begin{equation}
(2A\bar\psi\psi+3)W^{k}=-\frac{3}{4}\bar{\psi}\gamma^{k}\gamma_{5}\psi
+\nabla_{i}\left(C\bar\psi\gamma^5S^{ik}\psi
+iB\bar\psi\gamma^5\psi g^{ik}\right)
\end{equation}
and the matter field equations
\begin{equation}
i\Gamma^iD_i\psi-C\nabla_{i}W_{j}\gamma^5S^{ij}\psi-iB\nabla_{k}W^{k}\gamma^5\psi 
-AW^{k}W_{k}\psi-m\psi=0.
\end{equation}

The former equation can be inverted as to give the torsion in terms of the spinor fields and their derivatives, and then substituted into the latter equation. The resulting matter field equations contain second-order derivatives of the spinor that cannot be present in the Dirac field equations, much as in the previous case: to avoid such difficulty however, in this case it is enough to require $B=C=0$, and so the matter field equation is given by
\begin{equation}
i\Gamma^i\tilde{D}_i\psi
+\frac{9A}{16(2A\bar\psi\psi+3)^{2}}\bar{\psi}\gamma^{k}\psi\bar{\psi}\gamma_{k}\psi\psi
-\frac{9}{16(2A\bar\psi\psi+3)}\bar{\psi}\gamma^{k}\psi\gamma_{k}\psi-m\psi=0
\label{2}
\end{equation}
with no further derivatives of the spinor, and thus consistently defined and causal. It possesses a torsionally induced non-linear interaction given in terms of an undetermined coupling constant $A$ similarly to what happened in the case of minimal coupling discussed in \cite{Fabbri:2012yg}, although now the non-linearities are higher, and again such torsionally induced non-linear terms and the new interaction coming from the non-minimal coupling have a similar essence.

Performing the same analysis as above, following the Wilson criteria for renormalization by scaling the spinor according to the transformation $\sigma^{-\frac{3}{2}}$, we have that the matter field equation scales into
\begin{equation}
\begin{split}
&0=i\Gamma^i\tilde{D}_i\psi
+\frac{9A\sigma^{-5}}{16(2A\bar\psi\psi\sigma^{-3}+3)^{2}}\bar{\psi}\gamma^{k}\psi\bar{\psi}\gamma_{k}\psi\psi
-\frac{9\sigma^{-1}}{16(2A\bar\psi\psi\sigma^{-3}+3)}\bar{\psi}\gamma^{k}\psi\gamma_{k}\psi
-m\sigma^{2}\psi\approx \\
&\approx i\Gamma^i\tilde{D}_i\psi
-\sigma^{2}\left(-\frac{9}{64A|\bar\psi\psi|^{2}}\bar{\psi}\gamma^{k}\psi\bar{\psi}\gamma_{k}\psi
+\frac{9}{32A\bar\psi\psi}\bar{\psi}\gamma^{k}\psi\gamma_{k}+m\right)\psi\approx i\Gamma^i\tilde{D}_i\psi
\end{split}
\end{equation}
implying that whatever is the coupling constant of such self-interaction, its mass dimension must be negative, so that despite the fact that the non-renormalizability from which we started was higher than the one previously encountered, still the usual torsionally induced non-linear contribution and the new self-interaction given by the non-minimal coupling are, just like the mass term, not only renormalizable, but even super-renormalizable.

The results above give a second example in which the non-renormalizability due to the torsionally induced non-linear effects and the ones coming from the non-minimal coupling compensate leaving a renormalizable (and in fact super-renormalizable) matter field equation; in fact, it does so in a case with a higher degree of non-linearity. This pushes us to look for a third example, possibly with an even higher degree of non-linearity, or non-minimality in the coupling: this can be done with an interaction Lagrangian given by
\begin{equation}
\mathcal{L}_{\mathrm{interaction}}=k|\bar\psi\psi|^{b}W^{k}W_{k}\sqrt{|g|}
\end{equation}
in terms of the parameter $k$ and the index $b>1$ representing the degree of non-minimality of the coupling.

When all these terms are taken in the total action, variation with respect to torsion and spinor fields gives the torsion-spin field equations
\begin{equation}
(2k|\bar\psi\psi|^{b}+3)W^{k}=-\frac{3}{4}\bar{\psi}\gamma^{k}\gamma_{5}\psi
\end{equation}
and the matter field equations
\begin{equation}
i\Gamma^iD_i\psi-kbW^{k}W_{k}|\bar\psi\psi|^{b-1}\psi-m\psi=0
\end{equation}
as it can be checked; the former equation can be inverted as to give the torsion in terms of the spinor fields and then substituted into the latter equation yielding the explicit matter field equation
\begin{equation}
i\Gamma^i\tilde{D}_i\psi+\frac{9kb|\bar\psi\psi|^{b-1}}{16(2k|\bar\psi\psi|^{b}+3)^{2}}
\bar{\psi}\gamma^{k}\psi\bar{\psi}\gamma_{k}\psi\psi
-\frac{9}{16(2k|\bar\psi\psi|^{b}+3)}\bar{\psi}\gamma^{k}\psi\gamma_{k}\psi-m\psi=0
\label{3}
\end{equation}
possessing a torsionally induced non-linear interaction that is much worse, and again such torsionally induced non-linear terms and the new interaction coming from the non-minimal coupling give analogous contributions.

Nevertheless, Wilson criteria for renormalization can be applied to the matter field equation to see that
\begin{equation}
\begin{split}
&0=i\Gamma^i\tilde{D}_i\psi
+\frac{9kb|\bar\psi\psi|^{b-1}\sigma^{-3b-1}}{16(2k|\bar\psi\psi|^{b}\sigma^{-3b}+3)^{2}}
\bar{\psi}\gamma^{k}\psi\bar{\psi}\gamma_{k}\psi\psi
-\frac{9\sigma^{-1}}{16(2k|\bar\psi\psi|^{b}\sigma^{-3b}+3)}\bar{\psi}\gamma^{k}\psi\gamma_{k}\psi
-m\sigma^{2}\psi\approx\\
&\approx i\Gamma^i\tilde{D}_i\psi
-\sigma^{2}\left[\sigma^{3(b-1)}\left(-\frac{9b}{64k|\bar\psi\psi|^{b+1}}
\bar{\psi}\gamma^{k}\psi\bar{\psi}\gamma_{k}\psi
+\frac{9}{32k|\bar\psi\psi|^{b}}\bar{\psi}\gamma^{k}\psi\gamma_{k}\right)
+m\right]\psi\approx i\Gamma^i\tilde{D}_i\psi
\end{split}
\end{equation}
showing that the torsionally induced non-linear contribution and the self-interaction given by the non-minimal coupling are together such that the non-minimal terms go to zero even faster than the mass term, so that what might have been a worse non-minimal coupling was instead proven to be a much stronger renormalizability.

The idea is clear: our last mass dimensional analysis shows that in general, torsionally induced non-linear potentials in the matter field equations are such that if taken in non-minimal coupling they are not only renormalizable but also super-renormalizable, and the higher the non-minimal coupling becomes the stronger the super-renormalizability is.

This is an interesting result, and so far as we are aware, it depends on the presence of torsion and the non-minimal coupling, in the sense that no other theory of which we have any knowledge does that.

These results are about renormalizability of ultra-violet divergences, that is the worst we may encounter. It would however be better to check also that in the infra-red regime all equations work nicely. This can actually be done very quickly by considering the limit of weak fields given when the spinorial bilinear tend to zero: we have that in the three cases we had considered the matter field equations were given by \eqref{1}, \eqref{2} and \eqref{3}, or equivalently
\begin{equation}
i\Gamma^{h}\tilde{D}_{h}\psi-m\left[\frac{3}{16m(1+\epsilon\bar{\psi}\psi)}\left[(\bar{\psi}\psi)
+i(i\bar{\psi}\gamma_5\psi)\gamma_5\right]
+\left(\frac{\epsilon\bar\psi\psi-1}{\frac{1}{2}\epsilon\bar\psi\psi-1}\right)\right]\psi=0 
\end{equation}
and
\begin{equation}
i\Gamma^i\tilde{D}_i\psi
+\frac{9A(|\bar\psi\psi|^{2}+|i\bar\psi\gamma_5\psi|^{2})}{16(2A\bar\psi\psi+3)^{2}}\psi
-\frac{9}{16(2A\bar\psi\psi+3)}\left[(\bar{\psi}\psi)
+i(i\bar{\psi}\gamma_5\psi)\gamma_5\right]\psi-m\psi=0
\end{equation}
together with
\begin{equation}
i\Gamma^i\tilde{D}_i\psi
+\frac{9kb(|\bar\psi\psi|^{2}+|i\bar\psi\gamma_5\psi|^{2})|\bar\psi\psi|^{b-1}}
{16(2k|\bar\psi\psi|^{b}+3)^{2}}\psi
-\frac{9}{16(2k|\bar\psi\psi|^{b}+3)}\left[(\bar{\psi}\psi)
+i(i\bar{\psi}\gamma_5\psi)\gamma_5\right]\psi-m\psi=0
\end{equation}
after some Fierz rearrangement; in the infra-red regime we are allowed to take $i\bar{\psi}\gamma_5\psi=0$ so that we have 
\begin{equation}
i\Gamma^{h}\tilde{D}_{h}\psi-m\left[\frac{3\bar{\psi}\psi}{16m(1+\epsilon\bar{\psi}\psi)}
+\left(\frac{\epsilon\bar\psi\psi-1}{\frac{1}{2}\epsilon\bar\psi\psi-1}\right)\right]\psi=0 
\end{equation}
and
\begin{equation}
i\Gamma^i\tilde{D}_i\psi
+\frac{9A|\bar\psi\psi|^{2}}{16(2A\bar\psi\psi+3)^{2}}\psi
-\frac{9}{16(2A\bar\psi\psi+3)}\bar{\psi}\psi\psi-m\psi=0
\end{equation}
together with
\begin{equation}
i\Gamma^i\tilde{D}_i\psi
+\frac{9kb|\bar\psi\psi|^{b+1}}
{16(2k|\bar\psi\psi|^{b}+3)^{2}}\psi
-\frac{9}{16(2k|\bar\psi\psi|^{b}+3)}\bar{\psi}\psi\psi-m\psi=0
\end{equation}
and so for small $\bar\psi\psi$ they all reduce to the same
\begin{equation}
i\Gamma^i\tilde{D}_i\psi-\frac{3}{16}\bar{\psi}\psi\psi-m\psi=0
\end{equation}
which is the matter field equation we have in the minimal case, which we know to have no infra-red problem.

This concludes the survey about the properties of short-scale approximation as well as low-energy regimes, where we have shown that there are no problems for the matter field equations. On the other hand, there still are problems in the gravitational field equations, much like in the minimally coupled models. In fact we point out that for the gravitational field equations the problems of non-renormalizability are a little different, since the dynamical behaviour of the field equations (Einstein and Einstein--like equations) still comes from the fact that at short-distance ranges the kinetic term becomes irrelevant compared to the interacting term, not because the interacting term tends to become large, but because the kinetic term tend to become small. We will not deal with them in this paper because this suggests that the way out does not come from non-minimal coupling, but it has to be addressed within the framework of higher-order derivative theories. If we want no problem of renormalizability, we have to 
specifically focus on mass dimension $4$ theories \cite{s}. Thus, conformal gravity might be a possibility  \cite{Fabbri:2011vk,Fabbri:2011ha,Fabbri:2011jp,Fabbri:2012ws}.
\subsection{Parity Violation}
Now that we have settled the issue about renormalizability, let us try to investigate other phenomena related to these interactions, starting from the issue of parity violation: in the past, there has been some discussion about the possibility to allow parity violation in the gravitational action, so to have torsion inducing parity violation on the fermionic action as well \cite{Hojman:1980kv,Perez:2005pm,Alexandrov:2008iy}; we are not going to discuss here the implications about the Holst action and the Immirzi parameter, but merely we wish to point out that in the case of Dirac matter minimally coupled there remains no parity-oddness in the effective action \cite{Fabbri:2013gza}. Here we discuss what happens for Dirac matter non-minimally coupled.

To this extent, we return to study the case of mass dimension $5$ interaction, but now in parity odd-terms: take for example the interaction Lagrangian similar to the previous one, but now given in terms of the pseudo-scalar
\begin{equation}
\mathcal{L}_{\mathrm{interaction}}=(X\bar\psi\psi+Yi\bar\psi\gamma_5\psi)W^{k}W_{k}\sqrt{|g|}
\end{equation}
in terms of the generic parameters $X$ and $Y$.

Variation with respect to torsion and spinor fields gives the torsion-spin field equations
\begin{equation}
(2X\bar\psi\psi+2Yi\bar\psi\gamma_5\psi+3)W^{k}=-\frac{3}{4}\bar{\psi}\gamma^{k}\gamma_{5}\psi
\end{equation}
and the matter field equations
\begin{equation}
i\Gamma^iD_i\psi-(X+iY\gamma_5)W^{k}W_{k}\psi-m\psi=0.
\end{equation}
Inversion of the torsion-spin field equation and substitution into the matter field equations gives 
\begin{equation}
i\Gamma^i\tilde{D}_i\psi+\frac{9(|\bar\psi\psi|^{2}+|i\bar\psi\gamma_5\psi|^{2})}{16(2X\bar\psi\psi+2Yi\bar\psi\gamma_5\psi+3)^{2}}(X+iY\gamma_5)\psi
-\frac{9}{16(2X\bar\psi\psi+2Yi\bar\psi\gamma_5\psi+3)}
(\bar\psi\psi+i\bar\psi\gamma_5\psi i\gamma_5)\psi-m\psi=0
\end{equation}
which is not parity-even. The low-energy limit is such that $i\bar\psi\gamma_5\psi\approx0$ and then
\begin{equation}
i\Gamma^i\tilde{D}_i\psi+\left|\frac{3\bar\psi\psi}{4(2X\bar\psi\psi+3)}\right|^{2}
(X+iY\gamma_5)\psi-\frac{9\bar\psi\psi}{16(2X\bar\psi\psi+3)}\psi-m\psi\approx0
\end{equation}
still without a definite parity because of the term proportional to $Y\gamma_5$, and therefore it is only in the case in which the additional weak field approximation is taken that we have
\begin{equation}
i\Gamma^i\tilde{D}_i\psi-\frac{3\bar\psi\psi}{16}\psi-m\psi\approx0
\end{equation}
as above, and parity conservation is restored.

Again, the reason of parity violation is due to the non-minimal coupling.
\subsection{Additional Couplings}
To conclude this section, we would like to reconsider a somehow peculiar circumstance given by a non-minimal coupling that is nevertheless given for interactions of mass dimension $4$ (that is a coupling that is non-minimal) not because of the mass dimension but because of its non-standard structure: the mass dimension $4$ parity-even most complete interacting Lagrangian is given by
\begin{equation}
\mathcal{L}_{\mathrm{interaction}}=(pW^{\mu}\bar\psi\gamma_{\mu}\gamma_5\psi
+qT^{\mu}\bar\psi\gamma_{\mu}\psi)\sqrt{|g|}
\end{equation}
in terms of two constants $p$ and $q$ undetermined. We will employ the kinetic Lagrangian that has been given in \cite{Fabbri:2012yg} 
\begin{equation}\label{lagrangianTT}
\mathcal{L} = \left(R + A\/T_{ih}^{\;\;\;i}T^{jh}_{\;\;\;j} + B\/T_{ijh}T^{ijh} + C\/T_{ijh}T^{jih}\right)\sqrt{|g|}
\end{equation}
where $A$, $B$ and $C$ are coupling constants. When the total Lagrangian is considered, we may decompose the curvature in the torsionless curvature plus torsional terms, and further decompose these torsional terms into the three irreducible parts of torsion: when this is done, it is immediate to acknowledge that the non-completely antisymmetric irreducible part of torsion must vanish identically, and the two remaining vector parts of torsion are given by the system of field equations for torsion
\begin{eqnarray}
&W^{k}=\left(\frac{3}{8b}+\frac{p}{2b}\right)\bar{\psi}\gamma^{k}\gamma_{5}\psi\\
&T^{k}=\frac{q}{2a}\bar{\psi}\gamma^{k}\psi
\end{eqnarray}
where $a$ and $b$ are suitable combinations of the coupling parameters $A$, $B$ and $C$, while the field equation for the spinor field is expressed as
\begin{equation}
i\Gamma^i\tilde{D}_i\psi-\left[\frac{16bq^{2}-a(4p+3)^{2}}{32ab}\right]
\bar{\psi}\gamma^{\mu}\psi\gamma_{\mu}\psi-m\psi=0
\end{equation}
We notice that with respect to the minimal case $q=p=0$ here not only the coefficient of the coupling between dual-axial torsion and spinor pseudo-vector is shifted but additionally there is a new interaction between the trace torsion and the spinor vector, and in this sense the model is not minimally coupled; in the matter field equation, such non-minimal scheme only shifts the value of the constant in front of the interaction. The interacting potential can thus be made attractive or repulsive, weak or strong by simply tuning the four constants. In particular, the tuning $16bq^{2}-a(4p+3)^{2}=0$ even renders the matter field equation free.

By squaring the torsion-spin field equations and employing Fierz rearrangements, we get that 
\begin{eqnarray}
&-W^{2}\left(\frac{8b}{3+4p}\right)^{2}=T^{2}\frac{4a^{2}}{q^{2}}
=|\bar{\psi}\psi|^{2}+|i\bar{\psi}\gamma_5\psi|^{2}>0
\end{eqnarray}
showing that $W^{2}<0$ while $T^{2}>0$, and thus indicating that the dual-axial torsion has one physical component but the torsion trace has three physical components; the first result is as usual but the second result tells that there are three supplementary degrees of freedom. Therefore, we face here a peculiar circumstance, in which despite the fact that there appear to be three more degrees of freedom that take place in the dynamics, nevertheless all equations for all observational purposes are exactly like those one would have had in the minimally coupled counterpart.

All this seems to suggest that the degrees of freedom related to the torsion trace are not physical, or maybe they are real but dummy in the Dirac theory, due to the constrained structure of the Dirac spinor.
\section{Avoiding Cosmological Singularities}
As we have already mentioned, the issue of renormalizability of the Dirac equation has always been a sensible one since it has always been taken as one of the theoretical arguments against the presence of torsion; on the other hand, an issue that has always been considered as a theoretical argument in favour of torsion is the fact that the presence of torsion evade the singularity of the spacetime, as discussed in the already cited \cite{Poplawski:2009su,Fabbri:2011mg} and \cite{Poplawski:2011jz,Magueijo:2012ug}. In this paper, we have shown that torsion may still be present since there is no problem of non-renormalizability for the Dirac equation in the case of non-minimal coupling; on the other hand, it would just be ironic if the non-minimal coupling were to spoil the singularity avoidance that torsion allowed. Therefore, we should check that the presence of the non-minimal coupling does not create problems for the singularity avoidance torsion permitted.

The singularity avoidance in presence of torsion and fermions in minimal coupling is ensured by the fact that the Dirac equations do not admit singular solutions, as discussed in \cite{Poplawski:2009su,Fabbri:2011mg}; nevertheless, in these two papers the avoidance of singularity that is discussed is that of the Dirac particle itself, and therefore it may not necessarily apply to cosmological situations. The avoidance of singularity at a cosmological level must be studied independently as it has been done in \cite{Poplawski:2011jz,Magueijo:2012ug}, for instance. Other recent works are for instance \cite{Poplawski:2011wj,Poplawski:2010jv}, and \cite{Poplawski:2011qr}.

Here we would like to see that those results could be recovered also in the non-minimal coupling we are considering. To see that, let us begin by considering a Bianchi-I metric of the form
\begin{equation}
\label{4.1}
ds^2=dt^2-a^2(t)\,dx^2-b^2(t)\,dy^2-c^2(t)\,dz^2
\end{equation}
with tetrad fields
\begin{equation}
\label{4.2}
e^\mu_0=\delta^\mu_0, \quad e^\mu_1 = a(t) \delta^\mu_1, \quad e^\mu_2 = b(t) \delta^\mu_2, \quad e^\mu_3 = c(t) \delta^\mu_3
\end{equation}
and dual
\begin{equation}\label{4.3}
e^0_\mu = \delta^0_\mu, \quad e^1_\mu = \frac{1}{a(t)}\delta^1_\mu, \quad e^2_\mu = \frac{1}{b(t)}\delta^2_\mu, \quad e^3_\mu = \frac{1}{c(t)}\delta^3_\mu
\end{equation}
for $\mu =0,1,2,3$; the non-trivial coefficients of connection are
\begin{equation}\label{4.4}
\begin{split}
\tilde{\Gamma}_{10}^{\;\;\;1}= \frac{\dot a}{a}, \quad \tilde{\Gamma}_{20}^{\;\;\;2}= \frac{\dot b}{b}, \quad \tilde{\Gamma}_{30}^{\;\;\;3}= \frac{\dot c}{c}\\
\tilde{\Gamma}_{11}^{\;\;\;0}= a{\dot a}, \quad \tilde{\Gamma}_{22}^{\;\;\;0}= b{\dot b}, \quad \tilde{\Gamma}_{33}^{\;\;\;0}= c{\dot c}
\end{split}
\end{equation}
and in this case the matrices $\Gamma^i = e^i_\mu\gamma^\mu$ assume the explicit form
\begin{equation}\label{4.5}
\Gamma^0 = \gamma^0,\quad \Gamma^1 = \frac{1}{a(t)}\gamma^1, \quad \Gamma^2 = \frac{1}{b(t)}\gamma^2, \quad \Gamma^3 = \frac{1}{c(t)}\gamma^3 
\end{equation}
so that the spinorial-connection coefficients $\tilde{\Omega}_i$ are given by
\begin{equation}\label{4.8}
\tilde{\Omega}_0=0, \quad \tilde{\Omega}_1=\frac{1}{2}{\dot a}\gamma^1\gamma^0, \quad \tilde{\Omega}_2=\frac{1}{2}{\dot b}\gamma^2\gamma^0, \quad \tilde{\Omega}_3=\frac{1}{2}{\dot c}\gamma^3\gamma^0
\end{equation} 
and so the spinorial-covariant derivative induced by the Levi--Civita connection is
\begin{equation}\label{4.7}
\tilde{D}_i\psi = \partial_i\psi - \tilde{\Omega}_i\psi.
\end{equation}

Taking \eqref{4.8} and \eqref{4.7} into account, and defining $\tau=abc$, the gravitational and material equations assume the form
\begin{subequations}\label{4.10}
\begin{equation}\label{4.10a}
\begin{split}
\frac{\dot a}{a}\frac{\dot b}{b}+\frac{\dot b}{b}\frac{\dot c}{c}+\frac{\dot a}{a}\frac{\dot c}{c}=
\frac{1}{2\varphi}m\bar\psi\psi - \frac{3}{64\varphi^2}(\bar{\psi}\gamma_5\gamma^\tau\psi)(\bar{\psi}\gamma_5\gamma_\tau\psi)+\\
+\frac{1}{\varphi^2}\left[- \frac{3}{4}{\dot\varphi}^2 - \varphi\dot\varphi\frac{\dot\tau}{\tau}\right] - \frac{1}{2\varphi}V(\bar\psi\psi) 
\end{split}
\end{equation}
\begin{equation}\label{4.10b}
\begin{split}
\frac{\ddot b}{b} + \frac{\ddot c}{c} + \frac{\dot b}{b}\frac{\dot c}{c} = 
\frac{1}{\varphi^2}\left[\varphi\dot\varphi\frac{\dot a}{a} + \frac{3}{4}{\dot\varphi}^2 -\varphi\left( \ddot\varphi + \frac{\dot\tau}{\tau}\dot\varphi \right)\right] + \frac{3}{64\varphi^2}(\bar{\psi}\gamma_5\gamma^\tau\psi)(\bar{\psi}\gamma_5\gamma_\tau\psi)\\
- \frac{\epsilon(\bar\psi\psi)\left(\frac{m}{2}\bar\psi\psi -2V + \frac{3}{2}\bar\psi\psi V'\right)}{2\varphi\left(\frac{1}{2}\varphi - \frac{3}{2}\right)} - \frac{1}{2\varphi}V(\bar\psi\psi) + \frac{1}{2\varphi}(\bar\psi\psi)V'(\bar\psi\psi)
\end{split}
\end{equation}
\begin{equation}\label{4.10c}
\begin{split}
\frac{\ddot a}{a} + \frac{\ddot c}{c} + \frac{\dot a}{a}\frac{\dot c}{c} = 
\frac{1}{\varphi^2}\left[\varphi\dot\varphi\frac{\dot b}{b} + \frac{3}{4}{\dot\varphi}^2 -\varphi\left( \ddot\varphi + \frac{\dot\tau}{\tau}\dot\varphi \right) \right] + \frac{3}{64\varphi^2}(\bar{\psi}\gamma_5\gamma^\tau\psi)(\bar{\psi}\gamma_5\gamma_\tau\psi)\\
- \frac{\epsilon(\bar\psi\psi)\left(\frac{m}{2}\bar\psi\psi -2V + \frac{3}{2}\bar\psi\psi V'\right)}{2\varphi\left(\frac{1}{2}\varphi - \frac{3}{2}\right)} - \frac{1}{2\varphi}V(\bar\psi\psi) + \frac{1}{2\varphi}(\bar\psi\psi)V'(\bar\psi\psi)
\end{split}
\end{equation}
\begin{equation}\label{4.10d}
\begin{split}
\frac{\ddot a}{a} + \frac{\ddot b}{b} + \frac{\dot a}{a}\frac{\dot b}{b} = 
\frac{1}{\varphi^2}\left[\varphi\dot\varphi\frac{\dot c}{c} + \frac{3}{4}{\dot\varphi}^2 -\varphi\left( \ddot\varphi + \frac{\dot\tau}{\tau}\dot\varphi \right) \right] + \frac{3}{64\varphi^2}(\bar{\psi}\gamma_5\gamma^\tau\psi)(\bar{\psi}\gamma_5\gamma_\tau\psi)\\
- \frac{\epsilon(\bar\psi\psi)\left(\frac{m}{2}\bar\psi\psi -2V + \frac{3}{2}\bar\psi\psi V'\right)}{2\varphi\left(\frac{1}{2}\varphi - \frac{3}{2}\right)} - \frac{1}{2\varphi}V(\bar\psi\psi) + \frac{1}{2\varphi}(\bar\psi\psi)V'(\bar\psi\psi)
\end{split}
\end{equation}
\end{subequations}
and 
\begin{subequations}\label{4.9}
\begin{equation}\label{4.9a}
\dot\psi + \frac{\dot\tau}{2\tau}\psi + im\gamma^0\psi +
\frac{3i}{16\varphi} \left[ (\bar\psi\psi)\gamma^0 +i (i\bar\psi\gamma^5\psi)\gamma^0\gamma^5 \right]\psi + i\epsilon R\gamma^0\psi -iV'\gamma^0\psi=0
\end{equation}
\begin{equation}\label{4.9b}
\dot{\bar\psi} + \frac{\dot\tau}{2\tau}\bar\psi - im\bar{\psi}\gamma^0 - \frac{3i}{16\varphi}\bar\psi \left[ (\bar\psi\psi)\gamma^0 +i (i\bar\psi\gamma^5\psi)\gamma^5\gamma^0 \right] - i\epsilon R\bar\psi\gamma^0 + iV'\bar\psi\gamma^0=0
\end{equation}
\end{subequations}
together with the conditions
\begin{subequations}\label{4.11}
\begin{equation}\label{4.11a}
\tilde\Sigma_{12}=0\quad \Rightarrow \quad a \dot{b} - b \dot{a}=0 \quad \cup \quad \bar\psi\gamma^5\gamma^3\psi =0
\end{equation}
\begin{equation}\label{4.11b}
\tilde\Sigma_{23}=0\quad \Rightarrow \quad c \dot{b} - b \dot{c}=0 \quad \cup \quad \bar\psi\gamma^5\gamma^1\psi =0
\end{equation}
\begin{equation}\label{4.11c}
\tilde\Sigma_{13}=0\quad \Rightarrow \quad a \dot{c} - c \dot{a}=0 \quad \cup \quad \bar\psi\gamma^5\gamma^2\psi =0
\end{equation}
\end{subequations}
arising from the non-diagonal part of the gravitational equations (equations $\tilde\Sigma_{0A}=0$ ($A=1,2,3$) are automatically satisfied). There are three ways of satisfying these conditions: one is to impose constraints of purely geometrical origin by insisting that $a\dot{b}-b\dot{a}=0$, $a\dot{c}-c\dot{a}=0$, $c\dot{b}-b\dot{c}=0$ giving an isotropic universe filled with fermionic matter fields; another is to impose constraints of purely material origin by insisting that $\bar\psi\gamma^5\gamma^1\psi=0$, $\bar\psi\gamma^5\gamma^2\psi=0$, $\bar\psi\gamma^5\gamma^3\psi=0$ giving an anisotropic universe without fermionic torsional interactions; the third and last way is of both geometrical and material origin by insisting that $a\dot{b}-b\dot{a}=0$ with $\bar\psi\gamma^5\gamma^1\psi=0$, $\bar\psi\gamma^5\gamma^2\psi=0$ giving a partial isotropy.

For the remaining equations, following \cite{VFC}, we can suitably combine \eqref{4.10}, to obtain the equations
\begin{subequations}\label{4.12}
\begin{equation}\label{4.12a}
\varphi\tau\frac{d}{dt}\left(\frac{\dot a}{a} - \frac{\dot b}{b}\right) + \varphi\dot\tau\left(\frac{\dot a}{a} - \frac{\dot b}{b}\right) + \dot{\varphi}\tau\left(\frac{\dot a}{a} - \frac{\dot b}{b}\right)=0
\end{equation}
\begin{equation}\label{4.12b}
\varphi\tau\frac{d}{dt}\left(\frac{\dot a}{a} - \frac{\dot c}{c}\right) + \varphi\dot\tau\left(\frac{\dot a}{a} - \frac{\dot c}{c}\right) + \dot{\varphi}\tau\left(\frac{\dot a}{a} - \frac{\dot c}{c}\right)=0
\end{equation}
\end{subequations}
which can be directly integrated as
\begin{subequations}\label{4.13}
\begin{equation}\label{4.13a}
\frac{a}{b}=Ae^{\left(B\int{\frac{dt}{\varphi\tau}}\right)}
\end{equation}
\begin{equation}\label{4.13b}
\frac{a}{c}=Ce^{\left(D\int{\frac{dt}{\varphi\tau}}\right)}
\end{equation}
\end{subequations}
$A$, $B$, $C$ and $D$ being suitable constants; from \eqref{4.13} we get immediately
\begin{subequations}\label{4.13bis}
\begin{equation}\label{4.13bisa}
a= \tau^{\frac{1}{3}}\left(AC\right)^{\frac{1}{3}}e^{\left(\frac{B+D}{3}\int{\frac{dt}{\varphi\tau}}\right)}
\end{equation}
\begin{equation}\label{4.13bisb}
b=\tau^{\frac{1}{3}}A^{-\frac{2}{3}}C^{\frac{1}{3}}e^{\left(\frac{-2B+D}{3}\int{\frac{dt}{\varphi\tau}}\right)}
\end{equation}
\begin{equation}\label{4.13bisc}
c=\tau^{\frac{1}{3}}A^{\frac{1}{3}}C^{-\frac{2}{3}}e^{\left(\frac{B-2D}{3}\int{\frac{dt}{\varphi\tau}}\right)}
\end{equation}
\end{subequations}
and multiplying \eqref{4.10a} by 3, adding the result to the sum of \eqref{4.10b}, \eqref{4.10c} and \eqref{4.10d}, we get the dynamical equation
\begin{equation}\label{4.14}
2\frac{\ddot\tau}{\tau} = - 3\frac{\ddot\varphi}{\varphi} - 5\frac{\dot\tau}{\tau}\frac{\dot\varphi}{\varphi} - \frac{3m\bar\psi\psi 
-3\left(\varphi+1\right)V+3\varphi\bar\psi\psi V'}{\varphi\left(\varphi-3\right)}
\end{equation}
which can only be solved once a specific form of $V$ is given.

It is worth noticing that \eqref{4.10a} plays the role of a constraint on the initial data: thus for consistency we have to check that, if satisfied initially, this constraint is preserved in time. To see this point, we first observe that the Einstein-like equations \eqref{3.19}, and thus also \eqref{4.10}, can be written in the equivalent form
\begin{equation}\label{4.15}
\tilde{R}_{ij}= \tilde{T}_{ij} -\frac{1}{2}\tilde{T}g_{ij}
\end{equation}
where
\begin{equation}\label{4.16}
\begin{split}
\tilde{T}_{ij} = \frac{1}{\varphi}\tilde{\Sigma}_{ij}
+ \frac{1}{\varphi^2}\left( - \frac{3}{2}\de\varphi/de{x^i}\de\varphi/de{x^j} + \varphi\tilde{\nabla}_{j}\de\varphi/de{x^i} + 
\frac{3}{4}\de\varphi/de{x^h}\de\varphi/de{x^k}g^{hk}g_{ij}
-\varphi\tilde{\nabla}^h\de\varphi/de{x^h}g_{ij}\right) + \frac{3}{64\varphi^2}(\bar{\psi}\gamma_5\gamma^\tau\psi)(\bar{\psi}\gamma_5\gamma_\tau\psi)g_{ij} \\
- \frac{\epsilon(\bar\psi\psi)\left(\frac{m}{2}\bar\psi\psi -2V + \frac{3}{2}\bar\psi\psi V'\right)}{2\varphi\left(\frac{1}{2}\varphi - \frac{3}{2}\right)}\,g_{ij} - \frac{1}{2\varphi}V(\bar\psi\psi)\,g_{ij} + \frac{1}{2\varphi}(\bar\psi\psi)V'(\bar\psi\psi)\,g_{ij}
\end{split}
\end{equation}
denotes the effective stress-energy tensor appearing on the right hand side of \eqref{3.19}, while $\tilde T$ is its trace. It is then a straightforward matter to verify that \eqref{4.12} and \eqref{4.14} can be equivalently obtained by suitably combining the space-space equations of the set \eqref{4.15}; thus, we have that solving \eqref{4.12} and \eqref{4.14} amounts to solve all the space-space equations of the set \eqref{4.15}. In addition, the conservation laws automatically imply the vanishing of the four-divergence with respect to the Levi--Civita covariant derivative of the Einstein-like equations \eqref{4.A.12}. These two facts allow to apply a result by Bruhat (see \cite{yvonne4}, Theorem 4.1, pag. 150) which ensures that the constraint \eqref{4.10a} is actually satisfied for all time.

Also the Dirac equations \eqref{4.9} can be suitably combined, giving 
\begin{subequations}\label{4.17}
\begin{equation}\label{4.17a}
\frac{d}{dt}\left(\tau\bar\psi\psi\right) + \frac{3\tau}{8\varphi}\left(i\bar\psi\gamma^5\psi\right)\left(\bar\psi\gamma^5\gamma^0\psi\right)=0
\end{equation}
and
\begin{equation}\label{4.17b}
\begin{split}
\frac{d}{dt}\left(i\tau\bar\psi\gamma^5\psi\right) - \frac{3\tau}{8\varphi}\left(\bar\psi\psi\right)\left(\bar\psi\gamma^5\gamma^0\psi\right) - 2m\tau\left(\bar\psi\gamma^5\gamma^0\psi\right)
-2\epsilon R\tau\left(\bar\psi\gamma^5\gamma^0\psi\right)
+ 2V'\tau\left(\bar\psi\gamma^5\gamma^0\psi\right)=0
\end{split}
\end{equation}
and also
\begin{equation}
\begin{split}
\frac{d}{dt}\left(\tau\bar\psi\gamma^5\gamma^0\psi\right) + 2m\tau\left(i\bar\psi\gamma^5\psi\right)
+2\epsilon R\tau\left(i\bar\psi\gamma^5\psi\right) 
-2V'\tau\left(i\bar\psi\gamma^5\psi\right)=0
\end{split}
\end{equation}
\end{subequations}
altogether implying
\begin{equation}\label{4.18}
\left(\bar\psi\psi\right)^2 + \left(i\bar\psi\gamma^5\psi\right)^2 + \left(\bar\psi\gamma^5\gamma^0\psi\right)^2 = \frac{K^2}{\tau^2}
\end{equation}
where $K$ is a constant. 

As in \cite{VFC}, we may search for solutions of the Dirac equations such that $i\bar\psi\gamma^5\psi\!=\!\bar\psi\gamma^5\gamma^0\psi\!=\!0$, in such a way that $\bar\psi\psi\!=\!\frac{K}{\tau}$ and therefore in such a way that \eqref{4.14} reduces to a differential equations for the only unknown $\tau$. We notice that, in the standard representation for the spinor field $\bar\psi\!=\!(A^{\dagger},B^{\dagger})$, it is possible to take the non-relativistic approximation where the expression for the bi-linear scalar spinor reduces to $\bar\psi\psi\!=\!A^{\dagger}A-B^{\dagger}B\!\approx\!A^{\dagger}A\geqslant0$; because the volume of the universe is positive, this implies that $K\!=\!\tau\bar\psi\psi\geqslant0$ in such a limit, and since $K$ is a constant, then $K\geqslant0$ in general. This is very important for the following of the paper.

In this way, we may multiply \eqref{4.14} by $\varphi$ and taking into account that $\varphi=1+ \epsilon\bar\psi\psi$ and $\bar\psi\psi = \frac{K}{\tau}$ we obtain
\begin{equation}\label{ded2}
2\frac{\ddot\tau}{\tau}\varphi+3\ddot\varphi+5\frac{\dot\tau}{\tau}\dot\varphi
=\frac{3mK}{\tau\left(2-\frac{\epsilon K}{\tau}\right)} - \frac{3\left(\epsilon K + 2\tau\right)V}{\tau\left(2-\frac{\epsilon K}{\tau}\right)} + \frac{3\left(\epsilon\/K+\tau\right)KV'}{\tau^2\left(2-\frac{\epsilon\/K}{\tau}\right)}
\end{equation}
which, together with the identity $2\ddot\tau\varphi+3\tau\ddot\varphi+5\dot\tau\dot\varphi = \frac{d^2}{dt^2}\left(2\tau-\epsilon K\ln\tau\right)$ yields
\begin{equation}\label{ded2bis}
\frac{d}{dt}\left[\frac{d}{dt}\left(2\tau-\epsilon K\ln\tau\right)\right]^{2}
=6\left[mK-\left(\epsilon K + 2\tau\right)V +\frac{\left(\epsilon\/K + \tau\right)K}{\tau}V'\right]\dot{\tau}
\end{equation}
The complexity of equation \eqref{ded2bis} depends on the explicit form of the potential of self-interaction $V$, but in the special case in which $V$ vanishes, equation \eqref{ded2bis} simplifies considerably since it may be written as
\begin{equation}
\label{state}
\frac{d}{dt}\left[\frac{d}{dt}\left(2\tau-\epsilon K\ln\tau\right)\right]^{2}=6mK\dot{\tau}
\end{equation}
The latter can be integrated as
\begin{equation}\label{ded2tris}
\frac{d}{dt}(2\tau-\epsilon K\ln{\tau})\!=\!\sqrt{6mK\tau\!-\!A}
\end{equation}
yielding a first--order differential equation for $\tau$ with integration constant $A$. Assuming $A$ be negative, equation \eqref{ded2tris} can be integrated as
\begin{equation}
t\!+\!B\!=\!\frac{2\sqrt{|A|}}{3mK}\left(\sqrt{\frac{6mK}{|A|}\tau\!+\!1}\right)\!+\!
\frac{2\epsilon K}{\sqrt{|A|}}\mathrm{arctanh}\left(\sqrt{\frac{6mK}{|A|}\tau\!+\!1}\right)
\label{neg}
\end{equation}
but as it is also clear, $A$ negative (with of course $\tau$ positive) means that the argument of the $\mathrm{arctanh}$ is larger than one and thus such function is ill-defined. Therefore we are forced to assume $A\geq 0$: in the case $A >0$ the differential equation is integrated as
\begin{equation}
t\!+\!B\!=\!\frac{2\sqrt{A}}{3mK}\left(\sqrt{\frac{6mK}{A}\tau\!-\!1}\right)\!-\!
\frac{2\epsilon K}{\sqrt{A}}\arctan{\left(\sqrt{\frac{6mK}{A}\tau\!-\!1}\right)}
\label{pos}
\end{equation}
which is well-defined whenever the volume is larger than a given lower-bound $\tau_{0}\!\geqslant\!\frac{A}{6mK}$ and thus showing that, regardless the value of $B$, there is no way in which the minimal volume $\tau_{0}$ can be zero; if $A=0$, we get the solution
\begin{equation}
t\!+\!B\!=\!\frac{\sqrt{2}\left(\epsilon\/K+2\tau\right)}{\sqrt{3mK\tau}}
\label{zero}
\end{equation}
from which again we cannot have zero scale volume at a finite time. In all these cases then, singularities are avoided as in the case of spin fluids \cite{SpinFluid}. It is worth noticing that, according to the previous discussion, the avoidance of singularities would seem strictly due to the presence of the non-minimal coupling term present in (\ref{ded2bis}). In fact, if $\epsilon\!=\!0$ \eqref{neg}  would reduce to
\begin{equation}
t\!+\!B\!=\!\frac{2\sqrt{|A|}}{3mK}\left(\sqrt{\frac{6mK}{|A|}\tau\!+\!1}\right)
\end{equation}
allowing zero scale volume $\tau=0$ at the finite time $t\!=\!-B+\frac{2\sqrt{|A|}}{3mK}$. This would seem in contrast to the results presented in \cite{Poplawski:2011jz} where the author shows the absence of cosmological singularity also in the case of minimal coupling. In this regard, it should be pointed out that our analysis is based exclusively on the exact field equations and therefore it is of purely mathematical nature, while in \cite{Poplawski:2011jz} some physical assumptions are made (e.g. the stress energy tensor of the Dirac field is averaged to one of a prefect fluid, the relation between the square of the spin fluid and the number density of the fermions as well as the use  of the effective numbers of thermal degrees of freedom etc.) in order to get the stated results. In addition, there are characteristics the geometry of the space-time considered which are different in our case like e.g. the value of the spatial curvature parameter or the isotropy of the metric. Such differences make the comparison of our results with the ones of \cite{Poplawski:2011jz} not as straightforward as it might appear at first sight.

Another interesting aspect associated with the non--minimal coupling we are studying is that if there was a (cosmological) time interval in which the first term on the right hand side of equation \eqref{pos} is negligible with respect to the second one, then in such a time interval we would have an expansion of the universe according to $\tau\!\sim\!\left(\tan{t}\right)^{2}$, which could account for an accelerated behaviour possibly fitting inflationary scenarios. The above mentioned circumstance could be achieved for example by assigning initial data and then integration constants such that $\sqrt{A}/K$ is very small.

The model outlined above is therefore rather intriguing, because it can solve the problem of the cosmological singularity in quite elegant a way and simultaneously it can address the issue of inflationary scenarios. Unfortunately, the model with $V=0$ is unable to account for cosmic acceleration at late time. This is easily seen still considering equation \eqref{pos}, this time evaluated for large values of $\tau$, obtaining an expansion of the scale volume as $\tau\!\sim\!t^{2}$; due to equations \eqref{4.13bis}, this assures isotropization of space--time but under a Friedmann dynamical behaviour.

Being the fermionic non--minimal coupling alone insufficient to address the dark energy issue, to face this problem we should allow a potential to enter in the Lagrangian, therefore taking equation \eqref{ded2bis} into account with a given potential $V$. As an example, we consider the potential
\begin{equation}
\label{ded8}
V\left(\bar\psi\psi\right)=-\frac{1}{6\left(\epsilon\bar\psi\psi +1\right)}
\end{equation}
picked specifically to simplify the structure of equation \eqref{ded2bis} and render it easily integrable. Indeed, with the choice \eqref{ded8}, equation \eqref{ded2bis} can be integrated as
\begin{equation}\label{ded11}
\frac{\left(2\tau-\epsilon K\right)\dot\tau}{\tau} = \sqrt{6mK\tau + \tau^2 +2C}
\end{equation}
with $C$ denoting an integration constant. It is evident that if  $C$ is negative there exists automatically a strictly positive minimum value of the scale volume. Therefore, we discuss the case $C > 0$; in such a circumstance, a further integration yields
\begin{equation}
\label{ded13}
t+D=2\ln{\left(\sqrt{6mK\tau + \tau^2 + 2C}+3mK+\tau\right)}\!+\!
\frac{\epsilon K}{\sqrt{2C}}\ln\left(\frac{2C+3mK\tau+\sqrt{2C}\sqrt{6mK\tau + \tau^2 + 2C}}{\tau}\right) 
\end{equation}
From equation \eqref{ded13}, it follows that $\tau=0$ is possible only at infinite cosmological time, moreover for large values of $\tau$ we have an exponential expansion of the scale volume, ensuring that the scale factors of the metric tensor isotropize and undergo an accelerated expansion. For the sake of completeness, in the case $C=0$ we have
\begin{equation}\label{ded13bis}
t+D=\frac{\epsilon\left(6mK+\tau\right)}{3m\sqrt{6mK\tau+\tau^2}} + 4\ln\left(\sqrt{6mK+\tau}+\sqrt{\tau}\right)
\end{equation}
showing that the same qualitative results as for $C>0$ hold.

We have shown that fermionic non-minimal couplings possibly together with self-interacting potentials can be useful to face issues as inflation and late-time accelerated behaviour of the universe, without losing the results about the cosmological singularity, already established and existing in the literature. In the framework of the fermionic non--minimal coupling we have proposed, we will devote a forthcoming paper to a systematical analysis of cosmological models associated with different kinds of potentials V. 

\section{Conclusion}
In this paper, we have considered the basic field content for a background filled with Dirac matter, and we have relaxed the hypothesis of minimally coupled fields: we have mainly considered non-minimal couplings of the type $R\bar\psi\psi$ as a prototype, but we have also investigated other non-minimal interactions, in order to be as little dependent as possible on the specific kind of coupling, increasing the generality of the results; in the non-minimal couplings we have studied, we have essentially investigated mass dimension $5$ couplings, but eventually we have also considered specific situations in which non-minimal coupling was achieved for mass dimension $4$ couplings, and we have also seen that parity-violating gravitational terms could non-trivially be included in the action.

Our results spanned a variety of problems: first of all we have discussed how the spin and energy tensors are improved, but in these non-minimal couplings also the conservation laws are improved, and we have given not only their form but also demonstrated their validity; we did it in one specific example, since the exact structure of the extra terms are strongly model-dependent. 
Then, we have been addressing the fundamental issue related to the problem of renormalizability. 
We have seen that when we take both non-minimal coupling and torsion, the renormalizability of Dirac equation is not only restored, but it is also improved up to super-renormalizability. As an additional point of strength, we have seen that the larger the mass dimension of the non-minimal coupling the more super-renormalizable the effective coupling of the interactions within the matter field equations themselves. In addition, we have shown that cosmological singularity formation can be avoided by torsion also in the case of the non--minimal coupling we have considered, thus achieving results analogous to those already obtained for other minimally and non--minimally coupled theories; this too has been done in the specific case of the Bianchi-I universe, but again the arguments followed were relatively general.

We note that our analysis was focused on the non-minimal coupling, but it was not devoted to study higher-order derivative theories: thus, even if the renormalizability of the matter field equations came as an interesting surprise, that such renormalizability will not be extended to the gravitational field equations is at the same time unfortunate but expected. In fact, the non-renormalizability of the matter field equations is due to the fact that at short distances the kinetic term tends to become irrelevant because the effective interactions tend to become more relevant, and thus renormalizability can be regained by diminishing the scaling weight of such effective interactions, by changing the type of coupling. But the non-renormalizability of the gravitational field equations is due to the fact that at short distances the kinetic term tends to become irrelevant regardless the structure of the interactions, and thus renormalizability can be regained by changing the type of kinetic term.

This would imply having a different type of theory that would lie outside our aim, but we have suggested possible directions for enterprising such an extension.

\end{document}